

\newcount\mgnf\newcount\tipi\newcount\tipoformule
\newcount\aux\newcount\driver\newcount\cind\global\newcount\bz
\newcount\tipobib\newcount\stile\newcount\modif
\newcount\noteno\noteno=1\newcount\pos\newcount\nofig

\newdimen\stdindent\newdimen\bibskip
\newdimen\maxit\maxit=0pt\newdimen\figsize\newdimen\figglue
\newdimen\nhsize

\stile=0         
\tipobib=1       
\bz=0            
\cind=0          
\mgnf=0          
\tipoformule=0   
\aux=1           
\pos=1		 
\nofig=0	 


\ifnum\mgnf=0
   \magnification=\magstep0 
   \hsize=17truecm\vsize=24truecm\hoffset=-0.5truecm\voffset=-1.0truecm
   \figsize=\hsize\nhsize=\hsize
    \parindent=4.pt\stdindent=\parindent\fi
\ifnum\mgnf=1
   \magnification=\magstep1\hoffset=-1.0truecm
   \voffset=-1.0truecm\hsize=18truecm\vsize=24.truecm
   \baselineskip=14pt plus0.1pt minus0.1pt \parindent=6pt
   \lineskip=4pt\lineskiplimit=0.1pt      \parskip=0.1pt plus1pt
   \figsize=20truecm\figglue=-1truecm\nhsize=\hsize
   \stdindent=\parindent\fi
\ifnum\mgnf=2
   \magnification=\magstep2\hoffset=-1.5truecm
   \voffset=-1.0truecm\hsize=19truecm\vsize=24.truecm
   \baselineskip=14pt plus0.1pt minus0.1pt \parindent=6pt
   \lineskip=4pt\lineskiplimit=0.1pt\fi


\def\fine#1{}
\def\draft#1{\bz=1\ifnum\mgnf=1\baselineskip=22pt 
			      \else\baselineskip=16pt\fi
   \ifnum\stile=0\headline={\hfill DRAFT #1}\fi\raggedbottom
    \setbox150\vbox{\parindent=0pt\centerline{\bf Figures' captions}\*}
    \def\gnuins ##1 ##2 ##3{\gnuinsf {##1} {##2} {##3}}
    \def\gnuin ##1 ##2 ##3 ##4 ##5 ##6{\gnuinf {##1} {##2} {##3} 
		{##4} {##5} {##6}} 
    \def\eqfig##1##2##3##4##5##6{\eqfigf {##1} {##2} {##3} {##4} {##5} {##6}}
    \def\eqfigbis##1##2##3##4##5##6##7	
             {\eqfigbisf {##1} {##2} {##3} {##4} {##5} {##6} {##7}}
    \def\eqfigfor##1##2##3##4##5##6##7
             {\eqfigforf {##1} {##2} {##3} {##4} {##5} {##6} {##7}}
      \def\fine ##1{\vfill\eject\parindent=0pt
	 	\def\geq(####1){}
               \unvbox150\vfill\eject\raggedbottom
                \centerline{FIGURES}\unvbox149 ##1}}


\newcount\prau

\def\titolo#1{\setbox197\vbox{ 
\leftskip=0pt plus16em \rightskip=\leftskip
\spaceskip=.3333em \xspaceskip=.5em \parfillskip=0pt
\pretolerance=9999  \tolerance=9999
\hyphenpenalty=9999 \exhyphenpenalty=9999
\ftitolo #1}}
\def\abstract#1{\setbox198\vbox{
     \centerline{\vbox{\advance\hsize by -2cm \parindent=0pt\it Abstact: #1}}}}
\def\parole#1{\setbox195\hbox{
     \centerline{\vbox{\advance\hsize by -2cm \parindent=0pt Keywords: #1.}}}}
\def\autore#1#2{\setbox199\hbox{\unhbox199\ifnum\prau=0 #1%
\else, #1\fi\global\advance\prau by 1$^{\simbau}$}
     \setbox196\vbox {\advance\hsize by -\parindent\copy196
	\ottopunti\baselineskip=8pt$^{\simbau}${#2}\vfill}}
\def\prima{\unvbox197\vskip1truecm\centerline{\unhbox199}
     \footnote{}{\unvbox196}\vskip1truecm\unvbox198\vskip1truecm\copy195}
\def\simbau{\ifcase\prau
	\or \dagger \or \ddagger \or * \or \star \or \bullet\fi}


        \let\e=\varepsilon


{\count255=\time\divide\count255 by 60 \xdef\oramin{\number\count255}
        \multiply\count255 by-60\advance\count255 by\time
   \xdef\oramin{\oramin:\ifnum\count255<10 0\fi\the\count255}}
\def\ora{\oramin }

\def\data{\number\day/\ifcase\month\or gennaio \or febbraio \or marzo \or
aprile \or maggio \or giugno \or luglio \or agosto \or settembre
\or ottobre \or novembre \or dicembre \fi/\number\year;\ \ora}

\setbox200\hbox{$\scriptscriptstyle \data $}


\newcount\pgn \pgn=1
\newcount\firstpage

\def\foglio{\number\numsec:\number\pgn\global\advance\pgn by 1}
\def\foglioa{A\number\numsec:\number\pgn\global\advance\pgn by 1}

\def\pagina{\vfill\eject}
\def\ppagina{\ifodd\pageno\pagina\null\pagina\else\pagina\fi}
\def\ppaginan{\ifodd-\pageno\pagina\null\pagina\else\pagina\fi}

\def\setind{\firstpage=\pageno}
\def\setcap#1{\null\def\titlecap{#1}\global\firstpage=\pageno}
\def\titletesi{Indici critici per sistemi fermionici in una dimensione}

\ifnum\stile=1
  \def\pagenumbers{\headline={%
  \ifnum\pageno=\firstpage\hfil\else%
     \ifodd\pageno\hfill{\sc\titlecap}~~{\bf\folio}%
      \else{\bf\folio}~~{\sc\titletesi}\hfill\fi\fi}
  \footline={\ifnum\bz=0
                   \hfill\else\rlap{\hbox{\copy200}\ $\st[\foglio]$}\hfill\fi}}
  \def\pagenumbersind{\headline={%
  \ifnum\pageno=\firstpage\hfil\else%
    \ifodd\pageno\hfill{\rm\romannumeral-\pageno}%
     \else{\rm\romannumeral-\pageno}\hfill\fi\fi}
  \footline={\ifnum\bz=0
                   \hfill\else\rlap{\hbox{\copy200}\ $\st[\foglio]$}\hfill\fi}}
\else
  \def\pagenumbers{\headline={\hfill}
     \footline={\ifnum\bz=0\hfill\folio\hfill
                \else\rlap{\hbox{\copy200}\ $\st[\foglio]$}
		   \hfill\folio\hfill\fi}}
\fi

\pagenumbers

\def\numeropag#1{
   \ifnum #1<0 \romannumeral -#1\else \number #1\fi
   }


\global\newcount\numsec\global\newcount\numfor
\global\newcount\numfig\global\newcount\numpar
\global\newcount\numteo\global\newcount\numlem

\numfig=1\numsec=0

\gdef\profonditastruttura{\dp\strutbox}
\def\senondefinito#1{\expandafter\ifx\csname #1\endcsname\relax}
\def\SIA #1,#2,#3 {\senondefinito{#1#2}%
\expandafter\xdef\csname#1#2\endcsname{#3}\else%
\write16{???? ma #1,#2 e' gia' stato definito !!!!}\fi}
\def\etichetta(#1){(\veroparagrafo.\veraformula)
\SIA e,#1,(\veroparagrafo.\veraformula)
 \global\advance\numfor by 1
\write15{\string\FU (#1){\equ(#1)}}
\9{ \write16{ EQ \equ(#1) == #1  }}}
\def \FU(#1)#2{\SIA fu,#1,#2 }
\def\etichettaa(#1){(A\veroparagrafo.\veraformula)
 \SIA e,#1,(A\veroparagrafo.\veraformula)
 \global\advance\numfor by 1
\write15{\string\FU (#1){\equ(#1)}}
\9{ \write16{ EQ \equ(#1) == #1  }}}
\def \FU(#1)#2{\SIA fu,#1,#2 }
\def\tetichetta(#1){\veroparagrafo.\veroteorema
\SIA e,#1,{\veroparagrafo.\veroteorema}
\global\advance\numteo by1
\write15{\string\FU (#1){\equ(#1)}}%
\9{\write16{ EQ \equ(#1) == #1}}}
\def\tetichettaa(#1){A\veroparagrafo.\veroteorema
\SIA e,#1,{A\veroparagrafo.\veroteorema}
\global\advance\numteo by1
\write15{\string\FU (#1){\equ(#1)}}%
\9{\write16{ EQ \equ(#1) == #1}}}
\def\letichetta(#1){\veroparagrafo.\verolemma
\SIA e,#1,{\veroparagrafo.\verolemma}
\global\advance\numlem by1
\write15{\string\FU (#1){\equ(#1)}}%
\9{\write16{ EQ \equ(#1) == #1}}}
\def\getichetta(#1){{\bf Fig. \verafigura}:
 \SIA e,#1,{\verafigura}
 \global\advance\numfig by 1
\write15{\string\FU (#1){\equ(#1)}}
\9{ \write16{ Fig. \equ(#1) ha simbolo  #1  }}}

\def\veroparagrafo{\number\numsec}\def\veraformula{\number\numfor}
\def\verafigura{\number\numfig}\def\veroteorema{\number\numteo}
\def\verolemma{\number\numlem}

\def\geq(#1){\getichetta(#1)\galato(#1)}
\def\Eq(#1){\eqno{\etichetta(#1)\alato(#1)}}
\def\eq(#1){&\etichetta(#1)\alato(#1)}
\def\Eqa(#1){\eqno{\etichettaa(#1)\alato(#1)}}
\def\eqa(#1){&\etichettaa(#1)\alato(#1)}
\def\teq(#1){\tetichetta(#1)\talato(#1)}
\def\teqa(#1){\tetichettaa(#1)\talato(#1)}
\def\leq(#1){\letichetta(#1)\talato(#1)}

\def\Eqr{\eqno(\veroparagrafo.\veraformula)\advance\numfor by 1}
\def\eqr{&(\veroparagrafo.\veraformula)\advance\numfor by 1}
\def\Eqar{\eqno(A\veroparagrafo.\veraformula)\advance\numfor by 1}
\def\eqar{&(A\veroparagrafo.\veraformula)\advance\numfor by 1}

\def\eqv(#1){\senondefinito{fu#1}$\clubsuit#1$\write16{Manca #1 !}%
\else\csname fu#1\endcsname\fi}
\def\equ(#1){\senondefinito{e#1}\eqv(#1)\else\csname e#1\endcsname\fi}


\newdimen\gwidth

\def\commenta#1{\ifnum\bz=1\strut \vadjust{\kern-\profonditastruttura
 \vtop to \profonditastruttura{\baselineskip
 \profonditastruttura\vss
 \rlap{\kern\hsize\kern0.1truecm
  \vbox{\hsize=1.7truecm\raggedright\nota\noindent #1}}}}\fi}
\def\talato(#1){\ifnum\bz=1\strut \vadjust{\kern-\profonditastruttura
 \vtop to \profonditastruttura{\baselineskip
 \profonditastruttura\vss
 \rlap{\kern-1.2truecm{$\scriptstyle#1$}}}}\fi}
\def\alato(#1){\ifnum\bz=1
 {\vtop to \profonditastruttura{\baselineskip
 \profonditastruttura\vss
 \rlap{\kern-\hsize\kern-1.2truecm{$\scriptstyle #1$}}}}\fi}
\def\galato(#1){\ifnum\bz=1 \gwidth=\hsize 
 {\vtop to \profonditastruttura{\baselineskip
 \profonditastruttura\vss
 \rlap{\kern-\gwidth\kern-1.2truecm{$\scriptstyle#1$}}}}\fi}

\def\inputfig#1{\ifnum\nofig=0 \input \string#1\else\vskip1truecm
\centerline{Figura}\vskip1truecm\fi}


\def\magg{\ifnum\mgnf=0\magstep0\else\magstep1\fi}

\newskip\ttglue

\font\ftitolo=cmbx12 
\font\eighttt=cmtt8 \font\sevenit=cmti7  \font\sevensl=cmsl8
\font\sc=cmcsc10

\font\msytwww=msbm7 scaled\magstep1


\def\settepunti{\def\rm{\fam0\sevenrm}
\textfont0=\sevenrm \scriptfont0=\fiverm \scriptscriptfont0=\fiverm
\textfont1=\seveni \scriptfont1=\fivei   \scriptscriptfont1=\fivei
\textfont2=\sevensy \scriptfont2=\fivesy   \scriptscriptfont2=\fivesy
\textfont3=\tenex \scriptfont3=\tenex   \scriptscriptfont3=\tenex
\textfont\itfam=\sevenit  \def\it{\fam\itfam\sevenit}%
\textfont\slfam=\sevensl  \def\sl{\fam\slfam\sevensl}%
\textfont\ttfam=\eighttt  \def\tt{\fam\ttfam\eighttt}
\textfont\bffam=\sevenbf 
\scriptfont\bffam=\fivebf \scriptscriptfont\bffam=\fivebf  
\def\bf{\fam\bffam\sevenbf}%
\tt \ttglue=.5em plus.25em minus.15em
\setbox\strutbox=\hbox{\vrule height6.5pt depth1.5pt width0pt}%
\normalbaselineskip=8pt\let\sc=\fiverm \normalbaselines\rm}



\let\nota=\settepunti
\def\ottopunti{\settepunti}

\def\nnn{\hbox{\msytwww N}}


\font\tenmib=cmmib10
\font\sevenmib=cmmib10 scaled 800

\textfont5=\tenmib  \scriptfont5=\sevenmib  \scriptscriptfont5=\fivei

\mathchardef\aaa= "050B
\mathchardef\xxx= "0518
\mathchardef\oo = "0521
\mathchardef\Dp = "0540
\mathchardef\H  = "0548
\mathchardef\FFF= "0546
\mathchardef\ppp= "0570
\mathchardef\nnn= "0517

\newdimen\xshift \newdimen\xwidth \newdimen\yshift \newdimen\ywidth
\newdimen\laln

\ifnum\pos=0\def\midinsert{}\def\endinsert{}\fi

\def\ins#1#2#3{\nointerlineskip\vbox to0pt {\kern-#2 \hbox{\kern#1 #3}
\vss}}

\def\lineauno#1#2#3#4{\ifnum\mgnf=1\hsize=\figsize\fi
\xwidth=#1 \xshift=\hsize \advance\xshift 
by-\xwidth \divide\xshift by 2
\yshift=#2 \divide\yshift by 2
\parindent=0pt
\line{\hglue\xshift \vbox to #2{\hsize #1\vfil 
#3 \includegraphics{#40.ps}}
\hfill}}

\def\unafig#1#2#3#4#5#6{
\setbox99\vbox{\parindent=0pt\par\ifnum\mgnf=1\hglue \figglue\fi
\lineauno{#1}{#2}{#3}{#4}
\nobreak
\*\*
\didascalia{\hsize=\nhsize\geq(#6)#5}}}

\def\eqfig#1#2#3#4#5#6{
\unafig{#1}{#2}{#3}{#4}{#5}{#6}
\midinsert\unvbox99\endinsert
}

\def\eqfigf#1#2#3#4#5#6{
\unafig{#1}{#2}{#3}{#4}{#5}{#6}
\midinsert\unvbox99\endinsert
\setbox149\vbox{\unvbox149 \*\* \centerline{Fig. \equ(#6)} 
\nobreak
\*
\nobreak
\noindent\ifnum\mgnf=1\hglue \figglue\fi\lineauno{#1}{#2}{#3}{#4}}
\setbox150\vbox{\unvbox150 \parindent=0pt\*{\bf Fig. \equ(#6)}: #5\*}
}

\def\lineadue#1#2#3#4#5{\ifnum\mgnf=1\hsize=\figsize\fi
\xwidth=#1 \multiply\xwidth by 2 
\xshift=\hsize \advance\xshift 
by-\xwidth \divide\xshift by 3
\yshift=#2 \divide\yshift by 2
\ywidth=#2
\parindent=0pt
\line{\hfill
\vbox to \ywidth{\vfil #3 \includegraphics{#50.ps}}
\hskip\xshift%
\vbox to \ywidth{\vfil #4 \includegraphics{#51.ps}}}}

\def\duefig#1#2#3#4#5#6#7{
\setbox99\vbox{\parindent=0pt\par\ifnum\mgnf=1\hglue \figglue\fi
\lineadue{#1}{#2}{#3}{#4}{#5}
\nobreak
\*\*
\didascalia{\hsize=\nhsize\geq(#7)#6}}}

\def\eqfigbisf#1#2#3#4#5#6#7{
\duefig{#1}{#2}{#3}{#4}{#5}{#6}{#7}
\midinsert\unvbox99\endinsert
\setbox149\vbox{\unvbox149 \*\* \centerline{Fig. \equ(#7)} 
\nobreak
\*
\nobreak
\ifnum\mgnf=1\hglue \figglue\fi\lineadue{#1}{#2}{#3}{#4}{#5}}
\setbox150\vbox{\unvbox150 \parindent=0pt\*{\bf Fig. \equ(#7)}: #6\*}
}

\def\eqfigbis#1#2#3#4#5#6#7{
\duefig{#1}{#2}{#3}{#4}{#5}{#6}{#7}
\midinsert\unvbox99\endinsert
}

\def\dimenfor#1#2{\par\xwidth=#1 \multiply\xwidth by 2 
\xshift=\hsize \advance\xshift 
by-\xwidth \divide\xshift by 3
\divide\xwidth by 2 
\yshift=#2 
\ywidth=#2}

\def\eqfigfor#1#2#3#4#5#6#7{
\midinsert
\parindent=0pt
\hbox to \hsize{\hskip\xshift 
\hbox to \xwidth{\vbox to \ywidth{\vfil#1\includegraphics{#50.ps}}\hfill}%
\hskip\xshift%
\hbox to \xwidth{\vbox to \ywidth{\vfil#2\includegraphics{#51.ps}}\hfill}\hfill}
\nobreak
\line{\hglue\xshift 
\hbox to \xwidth{\vbox to \ywidth{\vfil #3 \includegraphics{#52.ps}}\hfill}%
\hglue\xshift
\hbox to \xwidth{\vbox to\ywidth {\vfil #4 \includegraphics{#53.ps}}\hfill}\hfill}
\nobreak
\*\*
\didascalia{\geq(#7)#6}
\endinsert}

\def\eqfigforf#1#2#3#4#5#6#7{
\midinsert
\parindent=0pt
\hbox to \hsize{\hskip\xshift 
\hbox to \xwidth{\vbox to \ywidth{\vfil#1\includegraphics{#50.ps}}\hfill}%
\hskip\xshift%
\hbox to \xwidth{\vbox to \ywidth{\vfil#2\includegraphics{#51.ps}}\hfill}\hfill}
\nobreak
\line{\hglue\xshift 
\hbox to \xwidth{\vbox to \ywidth{\vfil #3 \includegraphics{#52.ps}}\hfill}%
\hglue\xshift
\hbox to \xwidth{\vbox to\ywidth {\vfil #4 \includegraphics{#53.ps}}\hfill}\hfill}
\nobreak
\*\*
\didascalia{\geq(#7)#6}
\endinsert
\setbox149\vbox{\unvbox149\* \centerline{Fig. \equ(#7)} \nobreak\* \nobreak
\*
\vbox{\hbox to \hsize{\hskip\xshift 
\hbox to \xwidth{\vbox to \ywidth{\vfil#1\includegraphics{#50.ps}}\hfill}%
\hskip\xshift%
\hbox to \xwidth{\vbox to \ywidth{\vfil#2\includegraphics{#51.ps}}\hfill}\hfill}
\nobreak
\line{\hglue\xshift 
\hbox to \xwidth{\vbox to \ywidth{\vfil #3 \includegraphics{#52.ps}}\hfill}%
\hglue\xshift
\hbox to \xwidth{\vbox to\ywidth {\vfil #4 \includegraphics{#53.ps}}\hfill}\hfill}
}\hfill}
\setbox150\vbox{\unvbox150\parindent=0pt\*{\bf Fig. \equ(#7)}: #6\*}
}

\def\eqfigter#1#2#3#4#5#6#7{
\line{\hglue\xshift 
\vbox to \ywidth{\vfil #1 \includegraphics{#2.ps}}
\hglue30pt
\vbox to \ywidth{\vfil #3 \includegraphics{#4.ps}}\hfill}
\multiply\xshift by 3 \advance\xshift by \xwidth \divide\xshift by 2
\line{\hfill\hbox{#7}}
\line{\hglue\xshift 
\vbox to \ywidth{\vfil #5 \includegraphics{#6.ps}}}}

\def\squaresym{\hbox to 7pt{\hskip 1pt\includegraphics{square.ps}\hfill}}
\def\crosssym{\hbox to 7pt{\hskip 1pt\includegraphics{cross.ps}\hfill}}
\def\boxsym{\hbox to 7pt{\hskip 1pt\includegraphics{box.ps}\hfill}}
\def\circlesym{\hbox to 7pt{\hskip 1pt\includegraphics{circle.ps}\hfill}}


\def\7{\ifnum\modif=1\write13\else\write12\fi}
\def\8{\immediate\write13}


\def\gnuin #1 #2 #3 #4 #5 #6{\midinsert\vbox{\vbox to 260pt{
\hbox to 420pt{
\hbox to 200pt{\hfill\nota (a)\hfill}\hfill
\hbox to 200pt{\hfill\nota (b)\hfill}}
\vbox to 110pt{\vfill\hbox to 420pt{
\hbox to 200pt{\includegraphics{#1.ps}\hfill}\hfill
\hbox to 200pt{\includegraphics{#2.ps}\hfill}
}}\vfill
\hbox to 420pt{
\hbox to 200pt{\hfill\nota (c)\hfill}\hfill
\hbox to 200pt{\hfill\nota (d)\hfill}}
\vbox to 110pt{\vfill\hbox to 420pt{
\hbox to 200pt{\includegraphics{#3.ps}\hfill}\hfill
\hbox to 200pt{\includegraphics{#4.ps}\hfill}
}}\vfill}
\vskip0.25cm
\0\didascalia{\geq(#5): #6}}
\endinsert}

\def\gnuinf #1 #2 #3 #4 #5 #6{\midinsert\nointerlineskip\vbox to 260pt{
\hbox to 420pt{
\hbox to 200pt{\hfill\nota (a)\hfill}\hfill
\hbox to 200pt{\hfill\nota (b)\hfill}}
\vbox to 110pt{\vfill\hbox to 420pt{
\hbox to 200pt{\includegraphics{#1.ps}\hfill}\hfill
\hbox to 200pt{\includegraphics{#2.ps}\hfill}
}}\vfill
\hbox to 420pt{
\hbox to 200pt{\hfill\nota (c)\hfill}\hfill
\hbox to 200pt{\hfill\nota (d)\hfill}}
\vbox to 110pt{\vfill\hbox to 420pt{
\hbox to 200pt{\includegraphics{#3.ps}\hfill}\hfill
\hbox to 200pt{\includegraphics{#4.ps}\hfill}
}}\vfill}
\?
\0\didascalia{\geq(#5): #6}
\endinsert
\global\setbox150\vbox{\unvbox150 \*\*\0 Fig. \equ(#5): #6}
\global\setbox149\vbox{\unvbox149 \*\*
    \vbox{\centerline{Fig. \equ(#5)(a)} \nobreak
    \vbox to 200pt{\vfill\includegraphics{#1.ps_f}}}\*\*
    \vbox{\centerline{Fig. \equ(#5)(b)}\nobreak
    \vbox to 200pt{\vfill\includegraphics{#2.ps_f}}}\*\*
    \vbox{\centerline{Fig. \equ(#5)(c)}\nobreak
    \vbox to 200pt{\vfill\includegraphics{#3.ps_f}}}\*\*
    \vbox{\centerline{Fig. \equ(#5)(d)}\nobreak
    \vbox to 200pt{\vfill\includegraphics{#4.ps_f}}}
}}

\def\gnuins #1 #2 #3{\midinsert\nointerlineskip
\vbox{\line{\vbox to 220pt{\vfill
\includegraphics{#1.ps}}\hfill}
\*
\0\didascalia{\geq(#2): #3}}
\endinsert}

\def\gnuinsf #1 #2 #3{\midinsert\nointerlineskip
\vbox{\line{\vbox to 220pt{\vfill
\includegraphics{#1.ps}}\hfill}
\*
\0\didascalia{\geq(#2): #3}}\endinsert
\global\setbox150\vbox{\unvbox150 \*\*\0 Fig. \equ(#2): #3}
\global\setbox149\vbox{\unvbox149 \*\* 
    \vbox{\centerline{Fig. \equ(#2)}\nobreak
    \vbox to 220pt{\vfill\includegraphics{#1.ps_f}}}} 
}


\def\9#1{\ifnum\aux=1#1\else\relax\fi}
\let\numero=\number
\def\boh{\hbox{$\clubsuit$}\write16{Qualcosa di indefinito a pag. \the\pageno}}
\def\didascalia#1{\vbox{\nota\baselineskip=9truept\0#1\hfill}\vskip0.3truecm}
\def\frac#1#2{{#1\over #2}}
\def\V#1{\underline{#1}}
	
		\let\i=\infty
		
 	\let\0=\noindent
\def\guida{\leaders\hbox to 1em{\hss.\hss}\hfill}
\def\tende#1{\vtop{\ialign{##\crcr\rightarrowfill\crcr
              \noalign{\kern-1pt\nointerlineskip}
              \hglue3.pt${\scriptstyle #1}$\hglue3.pt\crcr}}}
\def\otto{{\kern-1.truept\leftarrow\kern-5.truept\to\kern-1.truept}}

\def\={{ \; \equiv \; }}		
\ifnum\mgnf=0
    \def\openone{\leavevmode\hbox{\ninerm 1\kern-3.3pt\tenrm1}}%
\fi
\ifnum\mgnf=1
     \def\openone{\leavevmode\hbox{\ninerm 1\kern-3.6pt\tenrm1}}%
\fi

\def\2{{1\over2}}

\def\igb{
    \mathop{\raise4.pt\hbox{\vrule height0.2pt depth0.2pt width6.pt}
    \kern0.3pt\kern-9pt\int}}

\def\st{\scriptscriptstyle}
\let\\=\noindent
\def\*{\vskip0.5truecm}
\def\?{\vskip0.75truecm}
\def\item#1{\vskip0.1truecm\parindent=0pt\par\setbox0=\hbox{#1}
     \hangindent\ifdim\wd0>0.6cm 0.6cm\else\wd0\fi\hangafter 1 #1 \parindent=\stdindent}

\newdimen\notesize\notesize=\hsize \advance \notesize by -\parindent
\def\annota#1#2{{\footnote{${}^#1$}{\vtop {\hsize=\notesize\settepunti\
\baselineskip=8pt\hglue -\parindent#2\vfill}}}}
\def\annotano#1{\annota{\number\noteno}{#1}\advance\noteno by 1}


\def\ie{\hbox{\sl i.e.\ }}
\def\eg{\hbox{\sl e.g.\ }}
\def\qed{\hfill\break\nobreak\vbox{\vglue.25truecm\line{\hfill\raise1pt 
          \hbox{\vrule height9pt width5pt depth0pt}}}\vglue.25truecm}


\def\gint(#1)(#2)(#3){{\cal D}#1^{#2}\,e^{(#1^{#2+},#3#1^{#2-})}}

  \def\V0{{\bf 0}}    \def\tt{{\bf t}}

\def\FF{{\cal F}}

\def\PP{{\cal P}}


\ifnum\cind=1
\def\prtindex#1{\immediate\write\indiceout{\string\parte{#1}{\the\pageno}}}
\def\capindex#1#2{\immediate\write\indiceout{\string\capitolo{#1}{#2}{\the\pageno}}}
\def\parindex#1#2{\immediate\write\indiceout{\string\paragrafo{#1}{#2}{\the\pageno}}}
\def\subindex#1#2{\immediate\write\indiceout{\string\sparagrafo{#1}{#2}{\the\pageno}}}
\def\appindex#1#2{\immediate\write\indiceout{\string\appendice{#1}{#2}{\the\pageno}}}
\def\paraindex#1#2{\immediate\write\indiceout{\string\paragrafoapp{#1}{#2}{\the\pageno}}}
\def\subaindex#1#2{\immediate\write\indiceout{\string\sparagrafoapp{#1}{#2}{\the\pageno}}}
\def\bibindex#1{\immediate\write\indiceout{\string\bibliografia{#1}{Bibliografia}{\the\pageno}}}
\def\preindex#1{\immediate\write\indiceout{\string\premessa{#1}{\the\pageno}}}
\else
\def\prtindex#1{}
\def\capindex#1#2{}
\def\parindex#1#2{}
\def\subindex#1#2{}
\def\appindex#1#2{}
\def\paraindex#1#2{}
\def\subaindex#1#2{}
\def\bibindex#1{}
\def\preindex#1{}
\fi

\def\leaderfill{\leaders\hbox to 1em{\hss . \hss} \hfill }


\newdimen\capsalto \capsalto=0pt
\newdimen\parsalto \parsalto=20pt
\newdimen\sparsalto \sparsalto=30pt
\newdimen\tratitoloepagina \tratitoloepagina=2\parsalto
\def\aboveparteskip{\bigskip \bigskip}
\def\belowparteskip{\medskip \medskip}
\def\abovecapitskip{\bigskip}
\def\belowcapitskip{\medskip}
\def\belowparskip{\smallskip}
%


\def\parte#1#2{
   \9{\immediate\write16
      {#1     pag.\numeropag{#2} }}
   \aboveparteskip 
   \noindent 
   {\ftitolo #1} 
   \hfill {\ftitolo \numeropag{#2}}\par
   \belowparteskip
   }


\def\premessa#1#2{
   \9{\immediate\write16
      {#1     pag.\numeropag{#2} }}
   \abovecapitskip 
   \noindent 
   {\it #1} 
   \hfill {\rm \numeropag{#2}}\par
   \belowcapitskip
   }


\def\bibliografia#1#2#3{
  \ifnum\stile=1
   \9{\immediate\write16
      {Bibliografia    pag.\numeropag{#3} }}
   \belowcapitskip
   \noindent 
   {\bf Bibliografia} 
   \hfill {\bf \numeropag{#3}}\par
  \else
    \paragrafo{#1}{References}{#3}
\fi
   }


\newdimen\newstrutboxheight
\newstrutboxheight=\baselineskip
\advance\newstrutboxheight by -\dp\strutbox
\newdimen\newstrutboxdepth
\newstrutboxdepth=\dp\strutbox
\newbox\newstrutbox
\setbox\newstrutbox = \hbox{\vrule 
   height \newstrutboxheight 
   width 0pt 
   depth \newstrutboxdepth 
   }
\def\newstrut
   {\relax \ifmmode \copy \newstrutbox \else \unhcopy \newstrutbox \fi}
%
%
\vfuzz=3.5pt
%
%
\newdimen\indexsize \indexsize=\hsize
\advance \indexsize by -\tratitoloepagina
\newdimen\dummy
\newbox\parnum
\newbox\parbody
\newbox\parpage
%

%

\def\mastercap#1#2#3#4#5{
   \9{\immediate\write16
      {Cap. #3:#4     pag.\numeropag{#5} }}
   \abovecapitskip
   \setbox\parnum=\hbox {\kern#1\newstrut{#2}
			{\bf Capitolo~\number#3.}~}
   \dummy=\indexsize
   \advance\indexsize by -\wd\parnum
   \setbox\parbody=\vbox {
      \hsize = \indexsize \noindent \newstrut 
      {\bf #4}
      \newstrut \hss}
   \indexsize=\dummy
   \setbox\parnum=\vbox to \ht\parbody {
      \box\parnum
      \vfill 
      }
   \setbox\parpage = \hbox to \tratitoloepagina {
      \hss {\bf \numeropag{#5}}}
   \noindent \box\parnum\box\parbody\box\parpage\par
   \belowcapitskip
   }
\def\capitolo#1#2#3{\mastercap{\capsalto}{}{#1}{#2}{#3}}
%


%
\def\masterapp#1#2#3#4#5{
   \9{\immediate\write16
      {App. #3:#4     pag.\numeropag{#5} }}
   \abovecapitskip
   \setbox\parnum=\hbox {\kern#1\newstrut{#2}
			{\bf Appendice~A\number#3:}~}
   \dummy=\indexsize
   \advance\indexsize by -\wd\parnum
   \setbox\parbody=\vbox {
      \hsize = \indexsize \noindent \newstrut 
      {\bf #4}
      \newstrut \hss}
   \indexsize=\dummy
   \setbox\parnum=\vbox to \ht\parbody {
      \box\parnum
      \vfill 
      }
   \setbox\parpage = \hbox to \tratitoloepagina {
      \hss {\bf \numeropag{#5}}}
   \noindent \box\parnum\box\parbody\box\parpage\par
   \belowcapitskip
   }
\def\appendice#1#2#3{\masterapp{\capsalto}{}{#1}{#2}{#3}}
%

%

\def\masterpar#1#2#3#4#5{
   \9{\immediate\write16
      {par. #3:#4     pag.\numeropag{#5} }}
   \setbox\parnum=\hbox {\kern#1\newstrut{#2}\number#3.~}
   \dummy=\indexsize
   \advance\indexsize by -\wd\parnum
   \setbox\parbody=\vbox {
      \hsize = \indexsize \noindent \newstrut 
      #4
      \newstrut \hss}
   \indexsize=\dummy
   \setbox\parnum=\vbox to \ht\parbody {
      \box\parnum
      \vfill 
      }
   \setbox\parpage = \hbox to \tratitoloepagina {
      \hss \numeropag{#5}}
   \noindent \box\parnum\box\parbody\box\parpage\par
   \belowparskip
   }
\def\paragrafo#1#2#3{\masterpar{\parsalto}{}{#1}{#2}{#3}}
\def\sparagrafo#1#2#3{\masterpar{\sparsalto}{}{#1}{#2}{#3}}
%


%
\def\masterpara#1#2#3#4#5{
   \9{\immediate\write16
      {par. #3:#4     pag.\numeropag{#5} }}
   \setbox\parnum=\hbox {\kern#1\newstrut{#2}A\number#3.~}
   \dummy=\indexsize
   \advance\indexsize by -\wd\parnum
   \setbox\parbody=\vbox {
      \hsize = \indexsize \noindent \newstrut 
      #4
      \newstrut \hss}
   \indexsize=\dummy
   \setbox\parnum=\vbox to \ht\parbody {
      \box\parnum
      \vfill 
      }
   \setbox\parpage = \hbox to \tratitoloepagina {
      \hss \numeropag{#5}}
   \noindent \box\parnum\box\parbody\box\parpage\par
   \belowparskip
   }
\def\paragrafoapp#1#2#3{\masterpara{\parsalto}{}{#1}{#2}{#3}}
\def\sparagrafoapp#1#2#3{\masterpara{\sparsalto}{}{#1}{#2}{#3}}
%


\ifnum\stile=1

\def\newcap#1{\setcap{#1}
\vskip2.truecm\advance\numsec by 1
\\{\ftitolo \numero\numsec. #1}
\capindex{\numero\numsec}{#1}
\vskip1.truecm\numfor=1\pgn=1\numpar=1\numteo=1\numlem=1
}

\def\newapp#1{\setcap{#1}
\vskip2.truecm\advance\numsec by 1
\\{\ftitolo A\numero\numsec. #1}
\appindex{A\numero\numsec}{#1}
\vskip1.truecm\numfor=1\pgn=1\numpar=1\numteo=1\numlem=1
}

\def\newpar#1{
\vskip1.truecm
\vbox{
\\{\bf \numero\numsec.\numero\numpar. #1}
\parindex{\numero\numsec.\numero\numpar}{#1}
\*{}}
\nobreak
\advance\numpar by 1
}

\def\newpara#1{
\vskip1.truecm
\vbox{
\\{\bf A\numero\numsec.\numero\numpar. #1}
\paraindex{\numero\numsec.\numero\numpar}{#1}
\*{}}
\nobreak
\advance\numpar by 1
}

\else

\def\newsec#1{\vskip1.truecm
\advance\numsec by 1
\vbox{
\\{\bf \numero\numsec. #1}
\parindex{\numero\numsec}{#1}
\*{}}\numfor=1\pgn=1\numpar=1\numteo=1\numlem=1
\nobreak
}

\def\newsubsect#1{
\vskip1.truecm
\vbox{
\\{\it \numero\numsec.\numero\numpar. #1}
\parindex{\numero\numsec.\numero\numpar}{#1}
\*{}}
\nobreak
\advance\numpar by 1
}

\def\newapp#1{\vskip1.truecm
\advance\numsec by 1
\vbox{
\\{\bf A\numero\numsec. #1}
\appindex{A\numero\numsec}{#1}
\*{}}\numfor=1\pgn=1\numpar=1\numteo=1\numlem=1
}

\def\biblio{\vskip1.truecm
\vbox{
\\{\bf References.}\*{}
\bibindex{{}}}\nobreak\makebiblio
}

\fi


\newread\indicein
\newwrite\indiceout

\def\faindice{
\openin\indicein=\jobname.ind
\ifeof\indicein\relax\else{
\ifnum\stile=1
  \pagenumbersind
  \pageno=-1
  \setind
  \null
  \vskip 2.truecm
  \\{\ftitolo Indice}
  \vskip 1.truecm
  \parskip = 0pt
  \input \jobname.ind
  \ppaginan
\else
\\{\bf Index}
\*{}
 \input \jobname.ind
\fi}\fi
\closein\indicein
\def\nomeindice{\jobname.ind}
\immediate\openout \indiceout = \nomeindice
}


\newwrite\bib
\immediate\openout\bib=\jobname.bib
\global\newcount\bibex
\bibex=0
\def\verabib{\number\bibex}

\ifnum\tipobib=0
\def\cita#1{\expandafter\ifx\csname c#1\endcsname\relax
\hbox{$\clubsuit$}#1\write16{Manca #1 !}%
\else\csname c#1\endcsname\fi}
\def\rife#1#2#3{\immediate\write\bib{\string\raf{#2}{#3}{#1}}
\immediate\write15{\string\C(#1){[#2]}}
\setbox199=\hbox{#2}\ifnum\maxit < \wd199 \maxit=\wd199\fi}
\else
\def\cita#1{%
\expandafter\ifx\csname d#1\endcsname\relax%
\expandafter\ifx\csname c#1\endcsname\relax%
\hbox{$\clubsuit$}#1\write16{Manca #1 !}%
\else\probib(ref. numero )(#1)%
\csname c#1\endcsname%
\fi\else\csname d#1\endcsname\fi}%
\def\rife#1#2#3{\immediate\write15{\string\Cp(#1){%
\string\immediate\string\write\string\bib{\string\string\string\raf%
{\string\verabib}{#3}{#1}}%
\string\Cn(#1){[\string\verabib]}%
\string\CCc(#1)%
}}}%
\fi

\def\Cn(#1)#2{\expandafter\xdef\csname d#1\endcsname{#2}}
\def\CCc(#1){\csname d#1\endcsname}
\def\probib(#1)(#2){\global\advance\bibex+1%
\9{\immediate\write16{#1\verabib => #2}}%
}

\def\C(#1)#2{\SIA c,#1,{#2}}
\def\Cp(#1)#2{\SIAnx c,#1,{#2}}

\def\SIAnx #1,#2,#3 {\senondefinito{#1#2}%
\expandafter\def\csname#1#2\endcsname{#3}\else%
\write16{???? ma #1,#2 e' gia' stato definito !!!!}\fi}

\bibskip=10truept
\def\hboxto{\hbox to}

\catcode`\{=12\catcode`\}=12
\catcode`\<=1\catcode`\>=2
\immediate\write\bib<
	\string\halign{\string\hboxto \string\maxit%
	{\string #\string\hfill}&%
        \string\vtop{\string\parindent=0pt\string\advance\string\hsize%
	by -1.55truecm%
	\string#\string\vskip \bibskip
	}\string\cr%
>
\catcode`\{=1\catcode`\}=2
\catcode`\<=12\catcode`\>=12

\def\raf#1#2#3{\ifnum \bz=0 [#1]&#2 \cr\else
\llap{${}_{\rm #3}$}[#1]&#2\cr\fi}

\newread\bibin

\catcode`\{=12\catcode`\}=12
\catcode`\<=1\catcode`\>=2
\def\chiudibib<
\catcode`\{=12\catcode`\}=12
\catcode`\<=1\catcode`\>=2
\immediate\write\bib<}>
\catcode`\{=1\catcode`\}=2
\catcode`\<=12\catcode`\>=12
>
\catcode`\{=1\catcode`\}=2
\catcode`\<=12\catcode`\>=12

\def\makebiblio{
\ifnum\tipobib=0
\advance \maxit by 10pt
\else
\maxit=1.truecm
\fi
\chiudibib
\immediate \closeout\bib
\openin\bibin=\jobname.bib
\ifeof\bibin\relax\else
\raggedbottom
\input \jobname.bib
\fi
}

\openin13=#1.aux \ifeof13 \relax \else
\input #1.aux \closein13\fi
\openin14=\jobname.aux \ifeof14 \relax \else
\input \jobname.aux \closein14 \fi
\immediate\openout15=\jobname.aux

\def\V#1{\underline{#1}}

\def\normalbaselines{\baselineskip=20pt\lineskip=3pt\lineskiplimit=3pt}

\titolo{Thermodynamic entropy production fluctuation in a two
dimensional shear flow model}

\autore{F. Bonetto}{Mathmatics Department, Rutgers University,
New Brunswick, NJ 08903, email: {\tt bonetto@math.rutgers.edu}.}

\autore{J.L. Lebowitz}{Mathematics and Physics Department, 
Rutgers University, New Brunswick, NJ 08903, 
\ifnum\mgnf=1\hfill\break\indent\fi email: 
{\tt lebowitz@math.rutgers.edu}.}

\abstract{We investigate  fluctuations in
the momentum flux across a surface perpendicular to the velocity
gradient in a stationary shear flow maintained by either thermostated
deterministic or by stochastic boundary conditions. In the
deterministic system the Gallavotti-Cohen (GC)relation for the
probability of large deviations, which holds for the phase space
volume contraction giving the Gibbs ensemble entropy production, never
seems to hold for the flux which gives the hydrodynamic entropy
production. In the stochastic case the GC relation is found to hold
for the total flux, as predicted by extensions of the GC theorem but
not for the flux across part of the surface.  The latter appear to
satisfy a modified GC relation. Similar results are obtained for the
heat flux in a steady state produced by stochastic boundaries at
different temperatures.}

\prima
\newsec{Introduction}

There has been much effort during the past decade to connect
statistical mechanics of stationary nonequilibrium states (SNS) with
the theory of dissipative dynamical systems \cita{EQ}. While most
results obtained so far via this approach are more of a mathematical
than physical interest there is one potential exception: the
Gallavotti-Cohen theorem \cita{GC} and its generalizations
\cita{K}\cita{LS}\cita{M}. The original GC theorem
was motivated by the numerical results of \cita{ECM2} on a
``thermostated'' dynamical system. The phase-space time evolution of 
such a system is given by an equation of the form

$$\dot X=\FF(X)\Eq(eqm)$$ 
with $\FF$ chosen to keep $X(t)$ confined to a compact surface
$\Sigma$ in the phase-space while forcing the system into a
nonequilibrium state.  The latter requires that $\FF$ be
non-hamiltonian with ${\rm div } \FF(X) =\sigma(X)\not=0$.

Using some very strong assumptions on the dynamical system
\equ(eqm), GC proved that in the SRB measure
\cita{SRB} describing the SNS of this system the probability distribution
$P_\tau(p)=\langle\delta(p-\pi_\tau(X))\rangle$ of

$$\sigma_\tau(X)={1\over \tau\langle \sigma\rangle}
\int_{-\tau/2}^{\tau/2}\sigma(X(t))dt\Eq(st)$$

satisfies the equality

$$\lim_{\tau\to\i} {1\over \tau\langle \sigma\rangle}
\ln{P_\tau(p)\over P_\tau(-p)}=p\,.\Eq(flu)$$
Here $\langle\cdot\rangle$ represents the average in the SNS.

The quantity $\langle \sigma\rangle$ is formally equal to the ``rate
of change'' of the Gibbs entropy in the SNS. More precisely, if we
start the system with a measure $\mu_0(dX)=\rho_0(X)dX$
where $dX$ is the Liouville measure restricted to the surface $\Sigma$
then, using the evolution \equ(eqm),

$$\dot S_G(t)=-{d\over dt}\int \rho_t \log\rho_t dX=\mu_t(\sigma)
\tende{t\to\infty}\langle\sigma\rangle\Eq(Gibbs)$$
where the existence of the limit will hold under the assumptions of
the GC theorem. Furthermore we will have $\langle\sigma\rangle<0$
implying that $S_G(t)\to -\i$ whenever the limiting state is not an
equilibrium one with zero currents \cita{Ru},\cita{Ge}.

Based on the relation \equ(Gibbs) $\sigma_\tau(X)$ is often called the
(normalized) ``average entropy production during a time interval
$\tau$'' in the SNS.  The identification of $\sigma_\infty$, the object
of the GC theorem, with entropy production was further strengthened by
the form of $\sigma(X)$ in many examples of bulk thermostated systems,
e.g.\ those considered by Moran and Hoover
\cita{MH} for electrical conduction and by \cita{ECM} for shear
flow. In those systems $\sigma(X)$ is given by an expression related
to the hydrodynamic entropy production and the validity of the GC
relation, eq.\equ(flu), was confirmed by numerical simulations,
despite the fact that the conditions of the GC theorem are not
satisfied there.

Such bulk thermostated systems are however very different from
realistic systems which are typically driven to SNS by inputs at their
boundaries: the motion in their interiors being governed by
Hamiltonian dynamics which do not produce any
phase space volume contraction. It is therefore important, for
comparison with real systems, to consider models of dynamical systems
in which the thermostats forcing the system into SNS operate only
near the boundaries.  Such deterministic models were introduced by
Chernov and Lebowitz \cita{CL} for shear flow and by Gallavotti
\cita{EQ}, and van Beijeren \cita{vB} for heat flux.  In this paper we
investigate the validity and possible generalizations  of the GC relation
for such boundary driven systems.  Before doing that we note that the
GC theorem has been extended to open systems in contact with infinite
thermal reservoirs which act on the system but are not changed by it. The
simplest modeling of such a situation is via stochastic transitions,
induced by the reservoirs, between different microstates of the system
induced by the reservoirs \cita{LS}. Thus, to model a system carrying a
heat current  and/or a momentum flux, one may use Maxwellian boundary
conditions. This means that a particle hitting the left (right) wall will
be reflected with a Maxwellian distribution of velocities corresponding
to temperatures $T_L$ ($T_R$) and mean velocities $u_L$ ($u_R$)
parallel to these walls.  For $T_L\not = T_R$ this will induce a SNS with
a heat flux while $u_L\not = u_R$ will (using periodic boundary
conditions in the flow directions) induce a SNS with a shear flow, see
\cita{ELM}\cita{BLR}.
 It is expected (hoped) that the deterministic (thermostated) and
stochastic kinds of boundary modeling will yield similar SNS of a
macroscopic system away from the boundaries. This is what happens in
equilibrium systems, at least when not in a phase transition region. 

Such an ``equivalence of ensembles'' is far from established for SNS. In
fact there is, at some level, a profound difference between thermostated
and stochastically modeled SNS as far as $S_G$ is concerned. As ready
noted the former have $S_G(t)\to-\i$, and $\dot
S_G(t)\to\langle\sigma\rangle<0$, while the latter have $S_G(t)\to\bar
S_G $, $\dot S_G(t)\to 0$. The origin of the difference lies in the
differences in the measures describing these SNS. Thermostated SNS are
described by an SRB measure which is singular with respect to the induced
Lebesgue measure $dX$, while the SNS of stochastically driven systems are
(this can be proven in some case and expected in general) described by
measures that are absolutely continuous with respect to $dX$
\cita{GLP}\cita{GIK}. This difference need not however mean much for a
macroscopic system, since quantities of physical interest are sums of
functions which depend only on a few variables. Their properties are
therefore determined by the reduced distribution functions which can be
expected to be absolutely continuous with respect to the local Lebesgue
measure, \ie expressible as densities, even when the full measure is
singular and fractal \cita{BKL}.  

Interestingly enough, it is possible
for the boundary driven pure heat flow case to model the thermostat in such
a way that the expression for $\sigma(X)$ appearing in the GC relation is
the same, up to terms whose average vanishes, for both the deterministic
and stochastic case, see \cita{EQ}.  It can be written as 
$$\sigma=\left({1\over T_L}-{1\over T_R}\right)J_Q+{dF(X)\over dt}\Eq(flo_a)$$
so that

$$\langle\sigma\rangle=\left({1\over T_L}-{1\over T_R}\right)\langle
J_Q\rangle\Eq(flo)$$ 
where $J_{Q}(X)$ is the heat flux through some surface
in the middle of the system and the average of the time derivative $dF/dt$
vanishes in the SNS. $\langle \sigma \rangle$ has the form of entropy
production in the left and right heat ``reservoirs'' which is also
 equal to the hydrodynamic entropy production in the SNS \cita{BLR}.

The situation is different for the thermostated boundary driven shear
flow case considered by \cita{CL}. The $\sigma(X)$ in \equ(st),
entering the GC theorem for that model is not clearly related to the
hydrodynamic entropy production. The latter corresponds now to a momentum
flux through the system, $J_M(X)$, for which a GC relation holds for
the stochastically driven system. The question is then whether there
is still enough equivalence between systems driven deterministically
and stochastically so that the GC relation for $J_M(X)$, derived for
the latter, also holds in the former. 

Another question which concerns both the heat conduction and the shear
case is whether the GC relation can be observed in real macroscopic
physical systems. More precisely, we know that in the linear regime
the GC relations imply an Onsager type reciprocity relations for the
transport coefficients \cita{G1}, \cita{LS} which hold not only globally but
also locally.  Our question is therefore the following: assuming that the GC
relation holds for some flux crossing a surface $S$ does it also
hold (in some form) for the flux through part of $S$. The
reason this is important for the applicability of the GC relation to
real system is that for values of $p$ for which $P_\tau(p)$ is of order unity, $P_\tau(-p)$, in eq. \equ(flu), which corresponds to
the flux $J$ going in the opposite direction from its usual one, \eg
the heat flowing from the cold to the hot reservoir, is so small in a
macroscopic system that the possibility of observing it is effectively
zero. A local flux reversal on the other hand may be quite observable:
an attempt in this direction was indeed made by Ciliberto and Laroche 
\cita{Ci}.

Here we describe numerical investigations of these questions for a
deterministic and stochastically driven shear flow SNS and for a
stochastic heat flow model.

\newsec{Description of the shear flow model}

\0{\bf Deterministic}:   The
system consists of $N$ unit mass particles contained in a $L\times M$
box with periodic boundary conditions in the $x$-direction and
reflecting boundaries on the walls perpendicular to the $y$-direction.
The dynamics in the bulk of the system is Hamiltonian with hard core
interactions (the particles have a radius $r$). When a particle
collides with the reflecting walls its outgoing speed is the same as
the incoming one while the direction of the velocity is chosen in a
way which simulates a moving boundary and creates a shear flow.  The
boundary transformation we consider is the same as in
\cita{CL}\cita{BCL}.  Let $\varphi$ and $\psi$ be the angle that the
incoming (outgoing) velocity forms with the positive $x$-direction if
the particle collides with the upper wall or with the negative
$x$-direction if the particle collides with the lower wall. In the
thermostated system the outgoing angle $\psi$ is given by a function
of the incoming angle $\varphi$: $\psi =f(\varphi )$. The function $f$
we choose is $f(\varphi )=(\pi+b)-\sqrt{(\pi+b)^2-\varphi(\varphi-2b)}$.
This is time-reversible, \ie $ \pi -f(\pi-f(\varphi))=\varphi$ see Fig.
\equ(figure1). 

\eqfig{180pt}{120pt}{\ins{70pt}{100pt}{$\phi$}
\ins{109pt}{97pt}{$\psi=f(\phi)$}
\ins{-10pt}{53pt}{$L$}
\ins{82pt}{-4pt}{$M$}
}{sys}{Schematic representation of the dynamics of
the system.}{figure1}

\medskip
\0{\bf Stochastic}: The dynamics in the interior of the system is the 
same as before while the outgoing velocity at the upper (lower) wall
is now chosen as if the particle was coming from a Maxwellian bath at
temperature $\beta^{-1}$ moving at velocity $v_0 (-v_0)$. More
precisely we assume that after a collision with the boundary the
particle emerges with a velocity that is randomly chosen from a
distribution:
$$\PP(v)=\frac{1}{Z}v_y\exp\left(\frac{\beta}{2}
\left(\left(v_x-\epsilon_yv_0\right)^2+v^2_y\right)\right)\Eq(max)$$
where $Z$ is a normalization constant, $v_0$ is the mean $x$-momentum of
the particle after a collision and $\epsilon_{y}$ is 1 if the
collision is with the upper wall and $-1$ if it is with lower wall.
\medskip

In \cita{BCL} we checked the validity of the GC relation for the phase
space contraction generated during the collisions of particles with
the deterministic thermostated boundary.  We divided the phase space
contraction into a contribution due to the lower boundary and one due
to the upper one.  We found that the GC relation was well verified for
the total phase space contraction $\sigma(X)$ but not for the partial
ones\nobreak\annotano{We observe that based on the proof of GC we have
no reason to expect that such a relation should hold for the partial
phase space contraction.  We tested it anyway since, as already noted,
the GC relation appears to hold in more general situations than those
covered by the GC theorem, e.g. for the total $\sigma$ here.}.

As already noted there is no apparent connection between the phase
space contraction and the hydrodynamic entropy production which is 
proportional to the flux of the
$x$-component of the momentum across the system. In \cita{CL} equality
between average phase space contraction rate and average
hydrodynamical entropy production rate was shown to hold to first order in
the shear in the limit in which the system becomes large (at constant
density), \ie macroscopic, but the shear rate goes to zero in such a
way as to maintain a constant total momentum transfer. This was done
under the assumption that before a collision with either wall the
particle velocities are distributed according to a Maxwellian.  The
equality between these averages was supported by numerical evidence.

In the present paper we further investigate the possible equivalence
between phase space contraction and entropy production. The momentum
flux is equal to the momenta carried by
particles crossing a line through the middle of the system plus the
exchange of momentum between two colliding particles when their center
are on opposite side of the line, see fig.\equ(figure2).

\eqfig{180pt}{120pt}{
\ins{-10pt}{53pt}{$L$}
\ins{82pt}{-4pt}{$M$}
}{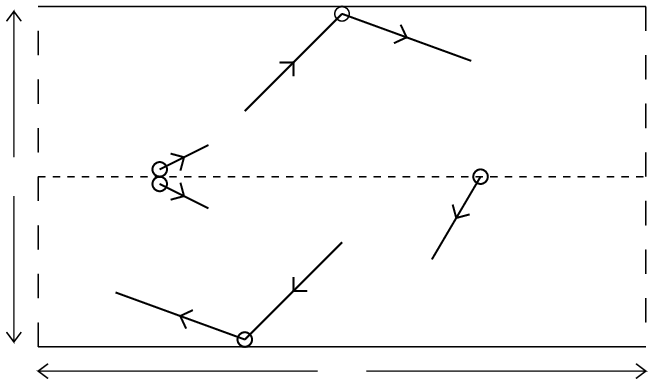}{Events producing momentum flux.}{figure2}

To be more precise we  first introduce a discrete time for the system (in a
slightly different way from what we did in \cita{BCL}).  Let
$X=(q_i,v_i)$ a phase space point, $\Phi_t(X)$ be the time evolution
induced by the dynamics and let $\tau (X)$ be the first time at which
a particle crosses the middle line or two particles on different sides
of this line collide starting from the phase point $X$ (we call such a
situation a {\it timing event}) and let
$S(X)=\Phi _{\tau (X)}(X)$.  Finally let $\pi (X)$ be the exchange
of $x$-momentum at the phase point $X$.

We now specify the quantity whose fluctuations we will check. Given an
integer $\tau$ let

$$
\pi _{\tau }(X)=\sum ^{\tau /2}_{i=-\tau /2}\pi (S^{i}(X))
\Eq(tau)$$
and

$$
p_{\tau }(X)=\frac{\pi _{\tau }(X)}{\left\langle \pi _{\tau }(X)\right\rangle }
\Eq(tau_m)$$
where the mean $\left\langle \cdot \right\rangle$ is taken with
respect to the stationary measure of the system.  Let now $\Pi _{\tau
}(p)$ be the distribution function of $p_{\tau }(X)$ and
$$\xi_\tau(p)=\frac{1}{\left\langle\pi_\tau(X)\right\rangle}
\ln \left(\frac{\Pi_\tau (p)}{\Pi_\tau(-p)}\right).$$
We can formulate our ``GC fluctuation relation for the entropy
production'' as follows:

$$\lim_{\tau\rightarrow\infty}\xi_\tau(p)=Cp
\Eq(Flu)$$
Here $C$ is the conversion constant between momentum flow and entropy
production that, from hydrodynamics \cita{BCL}, is $C=2u_b/T_b$ where
$u_b$ is the velocity of the particle near the upper boundary and $T_b$
is the temperature near the boundary.  Observe that this as well as
the following definitions, make sense also in the stochastic case when
the phase space contraction rate is not defined.
 
In this setting we define a local entropy production by looking at all
events like the ones described in figure \equ(figure2) that happen in
a specified part of the middle line of size $lL$.  More precisely let

$$\pi^l(X)=\pi (X)\chi _{[0,lL]}(X)\Eq(pi_p)$$
where $\chi _{A}(X)=1$ if the location of the momentum transfer event
is at a point in $A$ and 0 otherwise.  We now define in a way
analogous to eqs.\equ(tau) and \equ(tau_m) the quantities $\pi^l_\tau (X)$,
$p^l_\tau(X)$, $\Pi^l_\tau(p)$ and $\xi^l_\tau(p)$ and state our ``local
GC relation'' for the entropy production as,

$$\lim _{\tau \rightarrow\infty}\xi^l_\tau(p)=Cp\Eq(Flu_p).$$
More generally we check if a relation of the form:

$$\lim _{\tau \rightarrow\infty}\xi^l_\tau(p)=C_lp\Eq(Flu_pf)$$
holds.  

The stochastic model permits us to discuss also different kind of
transport phenomena. In fact we can set the reciprocal temperature
$\beta$ of the upper and lower walls to different value $\beta^+$ and
$\beta^-$. In this case we will have also transport of heat through
the middle of the system. We ran simulation also for this case,
setting $v_0=0$ for simplicity, and report the results in 
section 4. Clearly now the correct quantity to compute is the energy
transfered across the middle line when a particle crosses or
a collision happens.

\newsec{Numerical experiments for the shear flow}

We 
simulated systems with $ N=20$, $ 40$ and $ 60$ particles and $L,M$
such that the number density $ \delta =N/LM=0.034$ was fixed.  We
chose two different ways to set $L$ and $M$. In one case we fix $L=M$
\ie a square domain. In the second case we keep $M$ fixed at the value
it had for $N=20$ and increase $L$ proportionally as $N$ is
increased. In the deterministic case we fixed the energy per particle
${1\over 2N}\sum_iv_i^2=1$ while in the stochastic case we fixed the
values of $v_0$ and $\beta$ to reproduce the mean velocity $u_b$ and
temperature $T_b=\langle(v-u)^2\rangle$ observed in the $ N=60$
simulation for the deterministic system.  More precisely we fixed
$v_0=0.2$ and $\beta^{-1}=0.48$. Finally the radius $r$ was fixed to
1. For each value of $N$ and $L$ we followed a single trajectory of
the system for $5\cdot 10^{8}$ timing events and used it to compute
the distributions $ \Pi ^{l}_{\tau }(p)$ .  As in \cita{BCL}
(differently than in \cita{BGG}) we did not discard any events
between two consecutive segments of length $\tau$ .

\newsubsect{Stochastic case}

As discussed in the introduction (see also section 5.1 for further
discussion) we expect the fluctuation relation to hold for the total
momentum flux corresponding to the hydrodynamic entropy production.
This can indeed be seen in figure \equ(fig_s) in which
$\xi_\tau(p)$ is plotted for $\tau=300$ and $N=60$ for the rectangular
geometry. The dashed line represent the theoretical prediction
$\xi_\infty(p)=0.769p$. Similar results hold for $N=20$ and $N=40$.

\dimen2=25pt \advance\dimen2 by 285pt
\dimen3=447pt
\eqfig{\dimen3}{330pt}{
\ins{434pt}{0pt}{$\hbox to20pt{\hfill ${\scriptstyle p}$\hfill}$}
\ins{21pt}{310pt}{$\hbox to25pt{\hfill ${\scriptstyle \xi^l_\tau(p)}$\hfill}$}
\ins{8pt}{3pt}{$\hbox to 73pt{\hfill${\scriptstyle 0.0}$\hfill}$}
\ins{82pt}{3pt}{$\hbox to 73pt{\hfill${\scriptstyle 0.1}$\hfill}$}
\ins{155pt}{3pt}{$\hbox to 73pt{\hfill${\scriptstyle 0.2}$\hfill}$}
\ins{228pt}{3pt}{$\hbox to 73pt{\hfill${\scriptstyle 0.3}$\hfill}$}
\ins{302pt}{3pt}{$\hbox to 73pt{\hfill${\scriptstyle 0.4}$\hfill}$}
\ins{375pt}{3pt}{$\hbox to 73pt{\hfill${\scriptstyle 0.5}$\hfill}$}
\ins{0pt}{26pt}{$\hbox to 23pt{\hfill${\scriptstyle 0.0}$}$}
\ins{0pt}{84pt}{$\hbox to 23pt{\hfill${\scriptstyle 0.1}$}$}
\ins{0pt}{142pt}{$\hbox to 23pt{\hfill${\scriptstyle 0.2}$}$}
\ins{0pt}{199pt}{$\hbox to 23pt{\hfill${\scriptstyle 0.3}$}$}
\ins{0pt}{257pt}{$\hbox to 23pt{\hfill${\scriptstyle 0.4}$}$}
}{fig_s}{Fluctuation relation for the  momentum flux in the 
stocastic shear flow with rectangular geometry. The filled circles 
(\circlesym) represent 
the experimental value for the total momentum flux, 
with error bars, for $\tau=400$ and $N=60$ while the dashed line is the 
theoretical prediction. The other data represent the partial fluctuation 
relation for $l=0.9$ (\squaresym), $l=0.6$ (\boxsym) and $l=0.3$ (\crosssym).  
In all three cases $\tau=400$.}{fig_s}

The relation in the strong form given by eq.\equ(Flu_p) appears not to
hold for $l<1$ but eq.\equ(Flu_pf) seems to hold as one observes a
linear behavior of $\xi^l_\tau(p)$ always in Fig. \equ(fig_s).

The results for $\xi_\tau^l$ can be used to obtain the behavior of the
slope $C_l$ as a function of $l$. We report the result in
Fig. \equ(fig_slo_s) for the three value of $N$ and the two different
geometries we have used. We observe that in the case in which we keep
the box square $C_l$ decrease with $N$ and seems to reach a limit
different from 1 when $N$ grows. If we just increase the horizontal
side of the box keeping its height constant $C_l$ increases with $N$
and it is not clear from the data what, if any, limit is reached
when $N\to\infty$.

Instead of fixing $l$ we also tried fixing the length $l*L$ but found
nothing interesting.

\dimen2=25pt \advance\dimen2 by 174pt
\dimen3=228pt
\eqfigbis{\dimen3}{215pt}{
\ins{210pt}{0pt}{$\hbox to20pt{\hfill ${\scriptstyle l}$\hfill}$}
\ins{12pt}{199pt}{$\hbox to25pt{\hfill ${\scriptstyle C_l}$\hfill}$}
\ins{9pt}{3pt}{$\hbox to 34pt{\hfill${\scriptstyle 0.0}$\hfill}$}
\ins{43pt}{3pt}{$\hbox to 34pt{\hfill${\scriptstyle 0.2}$\hfill}$}
\ins{78pt}{3pt}{$\hbox to 34pt{\hfill${\scriptstyle 0.4}$\hfill}$}
\ins{113pt}{3pt}{$\hbox to 34pt{\hfill${\scriptstyle 0.6}$\hfill}$}
\ins{147pt}{3pt}{$\hbox to 34pt{\hfill${\scriptstyle 0.8}$\hfill}$}
\ins{182pt}{3pt}{$\hbox to 34pt{\hfill${\scriptstyle 1.0}$\hfill}$}
\ins{0pt}{21pt}{$\hbox to 14pt{\hfill${\scriptstyle 0.1}$}$}
\ins{0pt}{67pt}{$\hbox to 14pt{\hfill${\scriptstyle 0.3}$}$}
\ins{0pt}{112pt}{$\hbox to 14pt{\hfill${\scriptstyle 0.5}$}$}
\ins{0pt}{157pt}{$\hbox to 14pt{\hfill${\scriptstyle 0.7}$}$}
}{\ins{210pt}{0pt}{$\hbox to20pt{\hfill ${\scriptstyle l}$\hfill}$}
\ins{12pt}{199pt}{$\hbox to25pt{\hfill ${\scriptstyle C_l}$\hfill}$}
\ins{9pt}{3pt}{$\hbox to 34pt{\hfill${\scriptstyle 0.0}$\hfill}$}
\ins{43pt}{3pt}{$\hbox to 34pt{\hfill${\scriptstyle 0.2}$\hfill}$}
\ins{78pt}{3pt}{$\hbox to 34pt{\hfill${\scriptstyle 0.4}$\hfill}$}
\ins{113pt}{3pt}{$\hbox to 34pt{\hfill${\scriptstyle 0.6}$\hfill}$}
\ins{147pt}{3pt}{$\hbox to 34pt{\hfill${\scriptstyle 0.8}$\hfill}$}
\ins{182pt}{3pt}{$\hbox to 34pt{\hfill${\scriptstyle 1.0}$\hfill}$}
\ins{0pt}{21pt}{$\hbox to 14pt{\hfill${\scriptstyle 0.1}$}$}
\ins{0pt}{67pt}{$\hbox to 14pt{\hfill${\scriptstyle 0.3}$}$}
\ins{0pt}{112pt}{$\hbox to 14pt{\hfill${\scriptstyle 0.5}$}$}
\ins{0pt}{157pt}{$\hbox to 14pt{\hfill${\scriptstyle 0.7}$}$}
}{fig_slo_s}{Slope $C_l$ as a function of $l$ in the stochastic shear 
flow for $N=20$ (\circlesym), $N=40$ (\crosssym) and $N=60$ (\boxsym). The 
left figure is for the square geometry while the right one is for the 
rectangular geometry.}{fig_slo_s}

\newsubsect{Deterministic case}

The situation looks very different in the deterministic case. No
fluctuation relation seems to hold even when we consider the full
momentum transfer. The result are shown in Fig. \equ(fig_tot_d) where
$\xi_\tau(p)$ is plotted for $\tau=100,200$ and
$300$. The dashed line represents the value predicted by
eq. \equ(Flu). Although we still observe a linear behavior, the slope
appears to be increasing with $\tau$ so that no limit seems to be
reached.  We will attempt an explanation of this phenomenon in the
last section.

\dimen2=25pt \advance\dimen2 by 285pt
\dimen3=447pt
\eqfig{\dimen3}{330pt}{
\ins{434pt}{0pt}{$\hbox to20pt{\hfill ${\scriptstyle p}$\hfill}$}
\ins{21pt}{310pt}{$\hbox to25pt{\hfill ${\scriptstyle \xi_\tau(p)}$\hfill}$}
\ins{8pt}{3pt}{$\hbox to 73pt{\hfill${\scriptstyle 0.0}$\hfill}$}
\ins{82pt}{3pt}{$\hbox to 73pt{\hfill${\scriptstyle 0.1}$\hfill}$}
\ins{155pt}{3pt}{$\hbox to 73pt{\hfill${\scriptstyle 0.2}$\hfill}$}
\ins{228pt}{3pt}{$\hbox to 73pt{\hfill${\scriptstyle 0.3}$\hfill}$}
\ins{302pt}{3pt}{$\hbox to 73pt{\hfill${\scriptstyle 0.4}$\hfill}$}
\ins{375pt}{3pt}{$\hbox to 73pt{\hfill${\scriptstyle 0.5}$\hfill}$}
\ins{0pt}{26pt}{$\hbox to 23pt{\hfill${\scriptstyle 0.0}$}$}
\ins{0pt}{59pt}{$\hbox to 23pt{\hfill${\scriptstyle 0.1}$}$}
\ins{0pt}{91pt}{$\hbox to 23pt{\hfill${\scriptstyle 0.2}$}$}
\ins{0pt}{124pt}{$\hbox to 23pt{\hfill${\scriptstyle 0.3}$}$}
\ins{0pt}{156pt}{$\hbox to 23pt{\hfill${\scriptstyle 0.4}$}$}
\ins{0pt}{188pt}{$\hbox to 23pt{\hfill${\scriptstyle 0.5}$}$}
\ins{0pt}{221pt}{$\hbox to 23pt{\hfill${\scriptstyle 0.6}$}$}
\ins{0pt}{253pt}{$\hbox to 23pt{\hfill${\scriptstyle 0.7}$}$}
\ins{0pt}{286pt}{$\hbox to 23pt{\hfill${\scriptstyle 0.8}$}$}
}{fig_tot_d1}{Fluctuation relation in the deterministic 
shear flow in the rectangular geometry for the total momentum flux with 
$\tau=100$ (\circlesym), $\tau=200$ (\crosssym) and $\tau=300$ (\boxsym).}
{fig_tot_d}

We observe that a relation like eq. \equ(Flu_pf) seems to hold if we
look at the partial momentum flux. This is clearly shown in
Fig. \equ(fig_par_d) where the behavior of $\xi^l_\tau(p)$ for $l=0.6$
and several values of $\tau$ is plotted for $N=60$ in the rectangular
geometry.  As in the stochastic case we can look at the behavior of
$C_l$ as a function of $l$ for both geometries. The results
are plotted in Fig. \equ(fig_slo_d). As can be seen the slope depends
only very weakly on the size of the system, at least for the square
geometry. The only value for which $C_l$ seems to depend on the size
is for $l=0.9$.

\dimen2=25pt \advance\dimen2 by 285pt
\dimen3=447pt
\eqfig{\dimen3}{330pt}{
\ins{434pt}{0pt}{$\hbox to20pt{\hfill ${\scriptstyle p}$\hfill}$}
\ins{21pt}{310pt}{$\hbox to25pt{\hfill ${\scriptstyle \xi_\tau^l(p)}$\hfill}$}
\ins{20pt}{3pt}{$\hbox to 49pt{\hfill${\scriptstyle 0.0}$\hfill}$}
\ins{70pt}{3pt}{$\hbox to 49pt{\hfill${\scriptstyle 0.2}$\hfill}$}
\ins{120pt}{3pt}{$\hbox to 49pt{\hfill${\scriptstyle 0.4}$\hfill}$}
\ins{169pt}{3pt}{$\hbox to 49pt{\hfill${\scriptstyle 0.6}$\hfill}$}
\ins{219pt}{3pt}{$\hbox to 49pt{\hfill${\scriptstyle 0.8}$\hfill}$}
\ins{269pt}{3pt}{$\hbox to 49pt{\hfill${\scriptstyle 1.0}$\hfill}$}
\ins{319pt}{3pt}{$\hbox to 49pt{\hfill${\scriptstyle 1.2}$\hfill}$}
\ins{368pt}{3pt}{$\hbox to 49pt{\hfill${\scriptstyle 1.4}$\hfill}$}
\ins{0pt}{26pt}{$\hbox to 23pt{\hfill${\scriptstyle 0.0}$}$}
\ins{0pt}{61pt}{$\hbox to 23pt{\hfill${\scriptstyle 0.1}$}$}
\ins{0pt}{96pt}{$\hbox to 23pt{\hfill${\scriptstyle 0.2}$}$}
\ins{0pt}{130pt}{$\hbox to 23pt{\hfill${\scriptstyle 0.3}$}$}
\ins{0pt}{165pt}{$\hbox to 23pt{\hfill${\scriptstyle 0.4}$}$}
\ins{0pt}{199pt}{$\hbox to 23pt{\hfill${\scriptstyle 0.5}$}$}
\ins{0pt}{234pt}{$\hbox to 23pt{\hfill${\scriptstyle 0.6}$}$}
\ins{0pt}{268pt}{$\hbox to 23pt{\hfill${\scriptstyle 0.7}$}$}
}{fig_par_d1}{Approach to a limit of the right hand side of eq.\equ(Flu_pf) 
for the deterministic case. In this case $l=0.6$ while $\tau=100$ 
(\circlesym), 300 (\crosssym), 500 (\boxsym) and 700 (\squaresym).}{fig_par_d}

It is interesting to observe that if the fluctuation relation was true
we would expect to observe a slope $C_l=2u_b/T_b$ as discussed for the
stochastic system. In this situation, and mainly for the square
geometry, the value of $u_b$ varies significatively from $N=20$ to
$N=60$ while $C_l$ remain almost constant. This and the result for the
total momentum transfer suggests that the fluctuation of the phase space
contraction rate and those of the momentum flux  behave differently.

\dimen2=25pt \advance\dimen2 by 174pt
\dimen3=234pt
\eqfig{230pt}{215pt}{
\ins{210pt}{0pt}{$\hbox to20pt{\hfill ${\scriptstyle l}$\hfill}$}
\ins{21pt}{199pt}{$\hbox to25pt{\hfill ${\scriptstyle C_l}$\hfill}$}
\ins{16pt}{3pt}{$\hbox to 38pt{\hfill${\scriptstyle 0.0}$\hfill}$}
\ins{54pt}{3pt}{$\hbox to 38pt{\hfill${\scriptstyle 0.2}$\hfill}$}
\ins{93pt}{3pt}{$\hbox to 38pt{\hfill${\scriptstyle 0.4}$\hfill}$}
\ins{131pt}{3pt}{$\hbox to 38pt{\hfill${\scriptstyle 0.6}$\hfill}$}
\ins{170pt}{3pt}{$\hbox to 38pt{\hfill${\scriptstyle 0.8}$\hfill}$}
\ins{0pt}{21pt}{$\hbox to 23pt{\hfill${\scriptstyle 0.1}$}$}
\ins{0pt}{48pt}{$\hbox to 23pt{\hfill${\scriptstyle 0.3}$}$}
\ins{0pt}{74pt}{$\hbox to 23pt{\hfill${\scriptstyle 0.5}$}$}
\ins{0pt}{101pt}{$\hbox to 23pt{\hfill${\scriptstyle 0.7}$}$}
\ins{0pt}{127pt}{$\hbox to 23pt{\hfill${\scriptstyle 0.9}$}$}
\ins{0pt}{153pt}{$\hbox to 23pt{\hfill${\scriptstyle 1.1}$}$}
\ins{0pt}{180pt}{$\hbox to 23pt{\hfill${\scriptstyle 1.3}$}$}
}{fig_slo_d}{Slope $C_l$ as a function of $l$ in the deterministic shear 
flow for $N=20$ (\circlesym), $N=40$ (\crosssym) and $N=60$ (\boxsym) in the 
rectangular geometry. The graph of the square geometry in analogous. 
}{fig_slo_d}

\newsec{Heat flow}

The stochastic boundary condition permits us to study the case in
which the two walls are kept at different temperatures $T_+$ and
$T_-$. In this case the hydrodynamics entropy production is proportional to the heat current or energy flux from the upper wall of the system to the lower
one. The events contributing to an exchange of energy are the same as those
considered in Fig. \equ(figure2) but now we consider the kinetic
energy of a particle passing through the middle line or the exchange
of energy in a collision between two particle that are on different
side of the middle line.

Analogously to what we did in section 2 we define $\e(X)$ as the energy 
exchange for a point $X$ that is on the Poincar\'e section with

$$\e_\tau(X)=\sum ^{\tau /2}_{i=-\tau /2}\e (S^{i}(X))
\Eq(tau_e)$$
and

$$e_\tau(X)={\e_\tau(X)\over\left\langle \e_\tau(X)\right\rangle}
\Eq(tau_em)$$
Let now $E_\tau(e)$ be the distribution function of $e_\tau(X)$ and

$$\xi_\tau(e)=\frac{1}{\left\langle\e_\tau(X)\right\rangle}
\ln \left(\frac{E_\tau(e)}{E_\tau(-e)}\right)$$
As before we expect that

$$\lim_{\tau\to\infty}\xi_\tau(e)=Ce\Eq(flu_e)$$
where $C$ is the proper conversion constant between heat flow and
entropy production:

$$C=\frac{1}{T_+}-\frac{1}{T_-}$$

Similarly we 
define $\e_l(X)$, $e_l(X)$, $E^l_\tau(e)$ and $\xi^l_\tau(e)$ and
check whether 

$$\lim_{\tau\to\infty}\xi^l_\tau(e)=C_le\Eq(flu_ep)$$

\dimen2=25pt \advance\dimen2 by 285pt
\dimen3=455pt
\eqfig{\dimen3}{330pt}{
\ins{432pt}{0pt}{$\hbox to20pt{\hfill ${\scriptstyle e}$\hfill}$}
\ins{21pt}{310pt}{$\hbox to25pt{\hfill ${\scriptstyle E_\tau(e)}$\hfill}$}
\ins{14pt}{3pt}{$\hbox to 62pt{\hfill${\scriptstyle 0.0}$\hfill}$}
\ins{76pt}{3pt}{$\hbox to 62pt{\hfill${\scriptstyle 0.1}$\hfill}$}
\ins{139pt}{3pt}{$\hbox to 62pt{\hfill${\scriptstyle 0.2}$\hfill}$}
\ins{201pt}{3pt}{$\hbox to 62pt{\hfill${\scriptstyle 0.3}$\hfill}$}
\ins{264pt}{3pt}{$\hbox to 62pt{\hfill${\scriptstyle 0.4}$\hfill}$}
\ins{326pt}{3pt}{$\hbox to 62pt{\hfill${\scriptstyle 0.5}$\hfill}$}
\ins{388pt}{3pt}{$\hbox to 62pt{\hfill${\scriptstyle 0.6}$\hfill}$}
\ins{0pt}{26pt}{$\hbox to 23pt{\hfill${\scriptstyle 0.0}$}$}
\ins{0pt}{70pt}{$\hbox to 23pt{\hfill${\scriptstyle 0.1}$}$}
\ins{0pt}{113pt}{$\hbox to 23pt{\hfill${\scriptstyle 0.2}$}$}
\ins{0pt}{156pt}{$\hbox to 23pt{\hfill${\scriptstyle 0.3}$}$}
\ins{0pt}{199pt}{$\hbox to 23pt{\hfill${\scriptstyle 0.4}$}$}
\ins{0pt}{242pt}{$\hbox to 23pt{\hfill${\scriptstyle 0.5}$}$}
\ins{0pt}{286pt}{$\hbox to 23pt{\hfill${\scriptstyle 0.6}$}$}
}{fig_tot_c}{Fluctuation relation for the total energy flux in the 
heat flow model. The experimental value are plotted with error bar for 
$\tau=100$  (\circlesym), $\tau=200$ (\crosssym) and $\tau=300$ (\boxsym) and 
$N=20$. The dashed line is the theoretical prediction.}{fig_tot_c}

The numerical experiments are very similar to the ones described in
section 3 but we considered only stochastic boundary conditions with rectangular geometry. Finally we
fixed $T_+=0.7$, $T_-=0.4$ and all the other parameters as in the
shear flow case. The global fluctuation relation is 
shown in figure \equ(fig_tot_c).  There it can be observed that the
fluctuations are smaller 
 than the shear flow case. In fact for $\tau=300$,  when in the shear
flow case the limiting behavior was reached, we have just 2 points for
the fluctuation. We interpret the results as showing an  approach to the
expected limit. Similar result are obtained for $N=40,60$. The plots
for the $\xi_\tau^l(e)$, again, appear very linear, and their slopes $C_l$
are shown in figure
\equ(fig_slo_c). 

\dimen2=25pt \advance\dimen2 by 174pt
\dimen3=234pt
\eqfig{\dimen3}{215pt}{
\ins{210pt}{0pt}{$\hbox to20pt{\hfill ${\scriptstyle l}$\hfill}$}
\ins{21pt}{199pt}{$\hbox to25pt{\hfill ${\scriptstyle C_l}$\hfill}$}
\ins{26pt}{3pt}{$\hbox to 19pt{\hfill${\scriptstyle 0.0}$\hfill}$}
\ins{45pt}{3pt}{$\hbox to 19pt{\hfill${\scriptstyle 0.1}$\hfill}$}
\ins{64pt}{3pt}{$\hbox to 19pt{\hfill${\scriptstyle 0.2}$\hfill}$}
\ins{83pt}{3pt}{$\hbox to 19pt{\hfill${\scriptstyle 0.3}$\hfill}$}
\ins{102pt}{3pt}{$\hbox to 19pt{\hfill${\scriptstyle 0.4}$\hfill}$}
\ins{122pt}{3pt}{$\hbox to 19pt{\hfill${\scriptstyle 0.5}$\hfill}$}
\ins{141pt}{3pt}{$\hbox to 19pt{\hfill${\scriptstyle 0.6}$\hfill}$}
\ins{160pt}{3pt}{$\hbox to 19pt{\hfill${\scriptstyle 0.7}$\hfill}$}
\ins{179pt}{3pt}{$\hbox to 19pt{\hfill${\scriptstyle 0.8}$\hfill}$}
\ins{198pt}{3pt}{$\hbox to 19pt{\hfill${\scriptstyle 0.9}$\hfill}$}
\ins{0pt}{21pt}{$\hbox to 23pt{\hfill${\scriptstyle 0.2}$}$}
\ins{0pt}{48pt}{$\hbox to 23pt{\hfill${\scriptstyle 0.3}$}$}
\ins{0pt}{74pt}{$\hbox to 23pt{\hfill${\scriptstyle 0.4}$}$}
\ins{0pt}{101pt}{$\hbox to 23pt{\hfill${\scriptstyle 0.5}$}$}
\ins{0pt}{127pt}{$\hbox to 23pt{\hfill${\scriptstyle 0.6}$}$}
\ins{0pt}{153pt}{$\hbox to 23pt{\hfill${\scriptstyle 0.7}$}$}
\ins{0pt}{180pt}{$\hbox to 23pt{\hfill${\scriptstyle 0.8}$}$}
}{fig_slo_c1}{Slope $C_l$ as a function of $l$ in the heat
flow model for $N=20$ (\circlesym), $N=40$ (\crosssym) and $N=60$ (\boxsym).}
{fig_slo_c}

Similar comments as the one for the rectangular geometry shear flow
hold here.

\newsec{Conclusions}

We now try to summarize the results of our numerical experiments.

\newsubsect{Global fluctuation}

In the case of the stochastic boundary conditions the GC relation
appears to be satisfied for both shear flow and heat conduction.  This is 
not surprising. 
As in \cita{LS}, \cita{M}, \cita{BLR} we consider the
stationary probability $P_\tau(X(t))$ on the space of the trajectories of the
system from time $-\tau$ to time $\tau$.  (This can be assumed to be
unique, see for example \cita{GLP}, \cita{GL}.)  Define then 
the time reversal operation $I((q_i,v_i))=(q_i,-v_i)$
then if $X(t)$ is a possible trajectory so is $I(X(t))$ and we can
consider the measure $\bar P=I^*P$.  Then as in \cita{LS}, \cita{M} one finds

$${dP\over d\bar
P}(X)=\exp\left\{R(X(\tau))-R(X(-\tau))+
\int_{-\tau}^{\tau}\sigma(X(t))dt\right\}\Eq(Flup0)$$ 
for some appropriate function $\sigma(X)$.  Here the lhs represent a
Radon-Nykodyn derivative and $R(X)$ is a boundary term.  The \cita{GC}
relation then follows.  

It is easy to see that in the shear flow model
$\sigma_\tau=\int_{-\tau}^{\tau}\sigma(X)$ is proportional to the
momentum entering the system from the lower wall minus the momentum
leaving the system from the upper wall. Due to the conservation of
momentum in the bulk of the system the total momentum flux $\pi_\tau$
through the middle of the system is proportional to $\sigma_\tau$ plus
corrections due to the variation of momentum in the upper and lower
half of the system. We expect these two quantities to fluctuate much
less than the momentum flux, at least if the system is big enough so
that we can expect to have a GC fluctuation relation for $\pi_\tau$
\annotano{We observe however that differently from the analysis in
\cita{LS} we do not have an a priori bound on the fluctuation of the
total momentum of the system.}. Similar considerations hold for the
heat conduction model.

The deterministic case is far less clear. We know that the average
phase-space volume contraction rate is equal to the hydrodynamic
entropy production rate, only in some limiting situation, see
\cita{CL}. Our numerical results, together with those in \cita{BCL}
show that this equality does not extend to their fluctuations.  We
observe, first of all, that although the CG fluctuation relation
appears not to hold $\xi_\tau(p)$ is linear in $p$ so
that we can rewrite eq.\equ(Flu) as:

$$\xi_\tau(p)=C_\tau p\Eq(Flup)$$
The most striking effect appear to be the divergence of the slope
$C_\tau$. A possible explanation of this can be based on the assumption that
the distribution $\Pi_\tau(p)$ to be close to a Gaussian, see [BCL], so that

$$C_\tau={2\over S(\tau)}$$
where

$$S(\tau)=\frac{2}{\tau}\sum_{t=-\tau}^{\tau}D(t)-
\frac{2}{\tau^{2}}\sum_{t=-\tau}^{\tau}|t|D(t)\Eq(dec)$$
is the integral of the $\pi$-autocorrelation function 
$D(t)=\langle\pi(\Phi^t(\cdot))\pi(\cdot)\rangle-\langle\pi\rangle^2$.

Now when  $\sum_{t=-\tau}^{\tau}D(t)$ converges to a finite value the GC
relation
reduces to the usual Green-Kubo relation. On the other hand if we
assume that $\sum_{t=-\tau}^{\tau}D(t)$ approaches $0$ when
$\tau\to\i$ then $C_\tau$ has to diverge as $\tau^{-1}$, which is in
good agreement with our numerical data. We were not able to check more
directly this relation due to the lengthy simulations involved but we
hope to come back to this in future work.

We already noted that the quantity that
enters the GC theorem for our deterministic shear flow model, \ie the
phase space contraction rate due to collisions, appears to have little in
common with the hydrodynamic entropy production. This is different from models of deterministically boundary thermostated
heat conduction systems in which the phase space contraction rate
assume the form of an entropy production \cita{EQ} \cita{LS}, \cita{vB}.
We expect that it is possible to introduce a deterministic forcing
term for the shear flow acting only at, or near, the boundary and
producing a phase space contraction rate equal to the hydrodynamic
entropy production, see \cita{CL}.  This would make the GC relation
dependent on the form of the boundary forcing term.

\newsubsect{Local fluctuation}

As we already noted in \cita{BCL} the local fluctuation, in both the
stochastic and deterministic case, do not satisfy a fluctuation law in
the form of eq.\equ(Flu_p) but they appear to be in good agreement
with the more general relation eq.\equ(Flu_pf). This seems to be contrary
to the results obtained in \cita{GG} and in 
\cita{M}. 

To resolve this apparent contradiction we note that in 
\cita{GG} Gallavotti considers a chain of weakly interacting Anosov
dynamical systems\annotano{The paper by Maes deals with a rather more
general situation but similar arguments apply also there.}\ and takes
as a subsystem a finite piece of the chain. The phase space
contraction rate is an extensive quantity as in the bulk thermostated
systems.  Furthermore the correlations in the chain decay
exponentially both in space and in time.  In a system with these
characteristics one can prove that the fluctuation of the phase space
contraction rate due to the degree of freedom of the subsystem satisfy
a fluctuation relation with corrections proportional to the boundary
of the subsystem times the correlation length.

In our situation none of the above characteristics is present. First
of all we are not able to divide the degrees of freedom between the
subsystem and the rest of the system. Moreover the phase space volume
contraction is present only at the boundary of the system and the
subsystem does not include any portion of the boundary. Finally we
expect the correlation length in our system to be very long 
(potentially infinite), i.e. bigger that the size of the subsystem
considered. For these reason we do not expect the arguments in
\cita{GG} and \cita{M} to be applicable to our situation. We observe
however that our system is closer to the experimental situation
described in \cita{Ci} and to possible other experiments one can
imagine doing.

As in \cita{BCL} we do not have any real explanation for the apparent
validity of eq.\equ(Flu_pf). We think that the GC fluctuation relation
can be extended to a partial relation of the form eq.\equ(Flu_pf) in a
wide range of situations.  These include a 
system of particles under
the influence of an electric field and a Gaussian thermostat, see
\cita{BGG} and \cita{BDLR} for more details.  We think that for the 
asymmetric simple exclusion process, one should be able to find an analytic
justification of this behavior.
\*
\0{\bf Acknowledgements}
We are indebted to E.G.D. Cohen and G. Gallavotti for many helpful
discussion and suggestion. Research supported in part by NSF Grant
DMR-9813268, and Air Force Grant F49620-98-1-0207.
 
\rife{vB}{vB}{private communication.}

\rife{BCL}{BCL}{F. Bonetto, N.I. Chernov, J.L. Lebowitz, 
``(Global and Local) Fluctuation of Phase Space Contraction 
in Deterministic Stationary Non-Equilibrium, Chaos {\bf 8},  823--833 (1998).}

\rife{BDLR}{BDLR}{F. Bonetto, D. Daems, J.L. Lebowit, V. Ricci,
``Properties of Stationary 
Nonequilibrium States in the Thermostated Periodic Lorentz Gas III:  
The many colliding particles system'', in preparation.}

\rife{BGG}{BGG}{F. Bonetto, G. Gallavotti, P.L. Garrido, ``Chaotic
principle: an experimental test'', Physica D {\bf 105}, 226--252 (1997).}

\rife{BKL}{BKL}{F. Bonetto, A.J. Kupiainen, J.L. Lebowitz,
``Perturbation theory for coupled Arnold cat maps: absolute
continuity of marginal distribution'', preprint.}

\rife{BLR}{BLR}{F. Bonetto, J.L. Lebowitz, L. Rey-Bellet,``Fourier's Law: a 
Challenge for Theorists'',  Mathematical Physics 2000, Edited by A. Fokas, 
A. Grigoryan, T. Kibble and B. Zegarlinsky, Imperial College Press, 128-151 
(2000).}

\rife{Ci}{CiL}{S. Ciliberto, S. Laroche, ``An experimental verification of the
Gallavotti-Cohen fluctuation theorem'', Journal de Physique IV {\bf 8}, 
215-219 (1998).}

\rife{CL}{CL}{N.I. Chernov, J.L. Lebowitz, ``Stationary Nonequilibrium
States for Boundary Driven Hamiltonian Systems'' JSP {\bf 86},
953--990 (1997).}

\rife{ECM}{ECM1}{D.J. Evans, E.G.D. Cohen, G.P. Morriss, 
``Viscosity of a simple fluid from its maximal Lyapunov exponent'', Phys. 
Rev. A {\bf 42} 5990-5997 (1990)}

\rife{ECM2}{ECM2}{D.J. Evans, E.G.D. Cohen, G.P. Morriss, ``Probability
of Second Law Violations in Shearing Steady Flows'',
Phys. Rew. Letters {\bf 71}, 2401--2404 (1993).}

\rife{ELM}{ELM}{R. Esposito, J.L. Lebowitz, R. Marra, ``Hydrodynamic limit of 
the stationary Boltzmann equation in a slab'' Comm. Math. Phys. {\bf 160},  
49-80 (1994).}

\rife{EQ}{G1}{G. Gallavotti, ``New Methods in Nonequilibrium
Gases and Fluids'', Open System and Information Dynamics, Vol. 
{\bf 6}, 101-136 (1999) }

\rife{GG}{G2}{G. Gallavotti, ``A local fluctuation theorem'', Physica A 
{\bf 263}, 39-50 (1999).}

\rife{G1}{G3}{G. Gallavotti, ``Extension of the Onsager's reciprocity
to large fields and the chaotic hypothesis'', PRL {\bf 77}, 4434-4437 (1996).}

\rife{GC}{GC}{G. Gallavotti, E.G.D. Cohen, ``Dynamical ensemble in a
stationary states'' Jour. Stat. Phys {\bf 80}, 931--970 (1995).}

\rife{Ge}{Ge}{G. Gentile, ``Large deviation rule for Anosov flows'', 
Forum Math. {\bf 10}, 89--118 (1998).}

\rife{GIK}{GKI}{S. Goldstein, C. Kipnis and N. Ianiro, ``Stationary States 
for a System with Stochastic Boundary Conditions''
J. Stat. Phys. {\bf 41}, 915--930 (1985).}

\rife{GLP}{GLP}{S. Goldstein, J.L. Lebowitz and E. Presutti, 
``Mechanical System with Stochastic Boundaries''
Colloquia Mathematica Societati J\'anos Bolay {\bf 27} 403--419 (1979).}

\rife{GL}{GLR}{S. Goldstein, J.L. Lebowitz, K. Ravishankar, ``Approach to 
equilibrium in models of a system in contact with a heat bath'', 
J. Stat. Phys. {\bf 43}, 303--315 (1986).}

\rife{K}{K}{J. Kurchan, ``Fluctuation theorem for stochastic dynamics'', 
Jour. Phys. A {\bf 31}, 3719-3729 (1998).}

\rife{LS}{LS}{J.L. Lebowitz, H. Spohn, ``A Gallavotti-Cohen-type symmetry 
in the large deviation functional for stochastic dynamics'', 
J. Stat. Phys. {\bf 95}, 333-365 (1999).}

\rife{M}{M}{C. Maes, ``The fluctuation theorem as a Gibbs property'',
J. Stat. Phys. {\bf 95}, 367--392 (1999).  }

\rife{MH}{MH}{Moran Hoover, ``Diffusion in a periodic Lorentz gas'', 
J. Stat. Phys. {\bf 48},  709--726 (1987).}

\rife{Ru}{Ru1}{D. Ruelle, ``Positivity of entropy production in 
nonequilibrium statistical mechanics'' J. Stat. Phys. {\bf 85}, 
1--23 (1996).}

\rife{SRB}{Ru2}{D. Ruelle, ``Dynamical Systems Approach to Nonequilibrium
Statistical Mechanics: An Introduction'', IHES/Rutgers, Lecture Notes, 1997.}

\biblio


\end